\documentclass[letterpaper,pra, twocolumn,showpacs,superscriptaddress, nofootinbib]{revtex4-1}

\usepackage[T1]{fontenc}
\usepackage{color}
\usepackage[english]{babel}
\usepackage{pifont}
\usepackage{mathrsfs}
\usepackage{amsmath}
\usepackage{graphicx}
\usepackage[unicode=true,pdfusetitle,
 bookmarks=true,bookmarksnumbered=false,bookmarksopen=false,
 breaklinks=false,pdfborder={0 0 0},pdfborderstyle={},backref=false,colorlinks=true]
 {hyperref}
\hypersetup{
 urlcolor=blue, citecolor=blue}

\makeatletter

\pdfpageheight\paperheight
\pdfpagewidth\paperwidth

\usepackage{old-arrows}

\AtBeginDocument{
  
}

\makeatother

\begin{document}
\title{Photon statistics of quantum light on scattering from rotating ground
glass}
\author{Sheng-Wen Li}
\affiliation{Center for quantum technology research, School of Physics, Beijing
Institute of Technology, Beijing 100081, People\textquoteright s Republic
of China}
\affiliation{Texas A\&M University, College Station, Texas 77843, USA}
\author{Fu Li}
\affiliation{Texas A\&M University, College Station, Texas 77843, USA}
\author{Tao Peng}
\affiliation{Texas A\&M University, College Station, Texas 77843, USA}
\author{G. S. Agarwal}
\affiliation{Texas A\&M University, College Station, Texas 77843, USA}
\date{\today}
\begin{abstract}
When a laser beam passes through a rotating ground glass (RGG), the
scattered light exhibits thermal statistics. This is extensively used
in speckle imaging. This scattering process has not been addressed
in photon picture and is especially relevant if non-classical light
is scattered by the RGG. We develop the photon picture for the scattering
process using the Bose statistics for distributing $N$ photons in
$M$ pixels. We obtain analytical form for the \emph{P}-distribution
of the output field in terms of the \emph{P}-distribution of the input
field. In particular we obtain a general relation for the $n$-th
order correlation function of the scattered light, i.e., $g_{\text{out}}^{(n)}\simeq n!\,g_{\text{in}}^{(n)}$,
which holds for any order-$n$ and for arbitrary input states. This
result immediately recovers the classical transformation of coherent
light to pseudo-thermal light by RGG.
\end{abstract}
\maketitle

\section{Introduction}

Laser light usually carries a (super-)Poisson photon statistics \citep{scully_quantum_1966,scully_quantum_1967,scully_quantum_1997,agarwal_quantum_2012},
but when a laser beam passes through a rotating ground glass (RGG),
the scattered light exhibits a thermal statistics\footnote{ Throughout
this paper, we focus on the ``thermal light'' with a single frequency,
namely, only one optical mode is in the thermal state, and all the
other field modes are in the vacuum state.}. This is known as the
``pseudo-thermal'' light, which has been well verified in experiments
\citep{martienssen_coherence_1964,arecchi_measurement_1965,arecchi_1a4_1966,arecchi_high-order_1966,estes_scattering_1971,pearl_approximate_1969},
and widely used for ghost imaging \citep{gatti_ghost_2004,valencia_two-photon_2005,ferri_high-resolution_2005},
sub-wavelength imaging and lithography \citep{cao_two-photon_2010,li_beyond_2019},
as well as some fundamental studies \citep{peng_delayed-choice_2014,peng_poppers_2015,ihn_second-order_2017}. 

It is quite interesting that the input Poisson statistics can be changed
to be the thermal one in such a simple way. Usually this is understood
in classical theory \citep{goodman_statistical_2000,goodman_speckle_2010}:
the light field $E$ collected at the photon detector is the superposition
of the sub-fields $\tilde{E}_{i}$ propagated from different positions
of the RGG. Due to the rotation of the disc, the phases and amplitudes
of these sub-fields $\tilde{E}_{i}$ are varying with time randomly.
According to the central limit theorem, their summation $E$, as a
random variable, exhibits a Gaussian distribution {[}$P(E)\sim\exp(-\beta|E|^{2})${]},
and thus the intensity ($I\sim|E|^{2}$) exhibits a negative-exponential
distribution, i.e., the thermal distribution \citep{mandel_fluctuations_1958,mandel_fluctuations_1959,goodman_speckle_2010,goodman_statistical_2000}.

In principle, the central limit theorem requires the sub-sources $\tilde{E}_{i}$
should be infinitely many point sources. {Experimentalaly, the number
of sub-sources is limitted} by the granules within the light spot
on the ground glass; thus the pseudo-thermal light generated in experiments
shows some deviation from the ideal thermal distribution \citep{mehringer_photon_2018,bender_customizing_2018,bender_introducing_2019}.

Rigorously speaking, the continuous distribution $P(I)$ for the light
intensity is a classical treatment, but not exactly equivalent with
the distribution $P_{n}$ for the quantized photon numbers, especially
in the few-photon regime. Moreover, if the input light is a non-classical
state, the above classical interpretation does not apply, and clearly
the output light is generally no longer the thermal one.

In this sense, although the transformation of coherent light on a
diffusing glass plate has been extensively investigated \citep{arecchi_measurement_1965,arecchi_1a4_1966,arecchi_high-order_1966,pearl_approximate_1969,estes_scattering_1971,goodman_statistical_2000,goodman_speckle_2010},
and achieved wide applications in speckle imaging \citep{gatti_ghost_2004,valencia_two-photon_2005,ferri_high-resolution_2005,shapiro_physics_2012,bender_customizing_2018,bender_introducing_2019},
most time it was based on the above classical understanding rather
than a fully-quantized photon picture. In particular, if the input
light intensity is at single photon level, or is scattered by very
few diffusers, the validity of above classical theory needs to be
checked more carefully. 

In this paper, we develop a quantum framework to study the full photon
number statistics of such light scattered from general input states.
The basic picture is, when $\mathtt{N}_{\text{\textsc{p}}}$ photons
are scattered randomly by the RGG, the photon number received by a
small area fluctuates stochastically. Therefore, the photon number
statistics is obtained by counting the combinations how these photons
are randomly distributed on the scattered light pattern, and that
naturally gives the thermal statistics.

Then for arbitrary input states, the scattered photon statistics also
can be well obtained by taking the input statistics into account.
Moreover, we prove that the scattered $g_{\text{out}}^{(2)}$ is always
twice of the input $g_{\text{out}}^{(2)}\simeq2g_{\text{in}}^{(2)}$
for any input states, including non-classical states. Besides, we
also obtain the \emph{P}-function of the scattered light in the many-diffuser
limit, and that further gives the relation $g_{\text{out}}^{(n)}\simeq n!\,g_{\text{in}}^{(n)}$
for high-order correlations of any order-$n$ and any input.

The paper is arranged as follows. In Sec.\,II, we discuss the basic
generation mechanism of the pseudo-thermal light. In Sec.\,III, we
obtain the scattered $g^{(2)}$-correlation for arbitrary input state.
In Sec.\,IV, we discuss the scattered \emph{P}-function and high-order
correlations. The summary is drawn in Sec.\,V.

\section{The pseudo-thermal light generation from photon scattering}

\begin{figure}
\includegraphics[width=0.85\columnwidth]{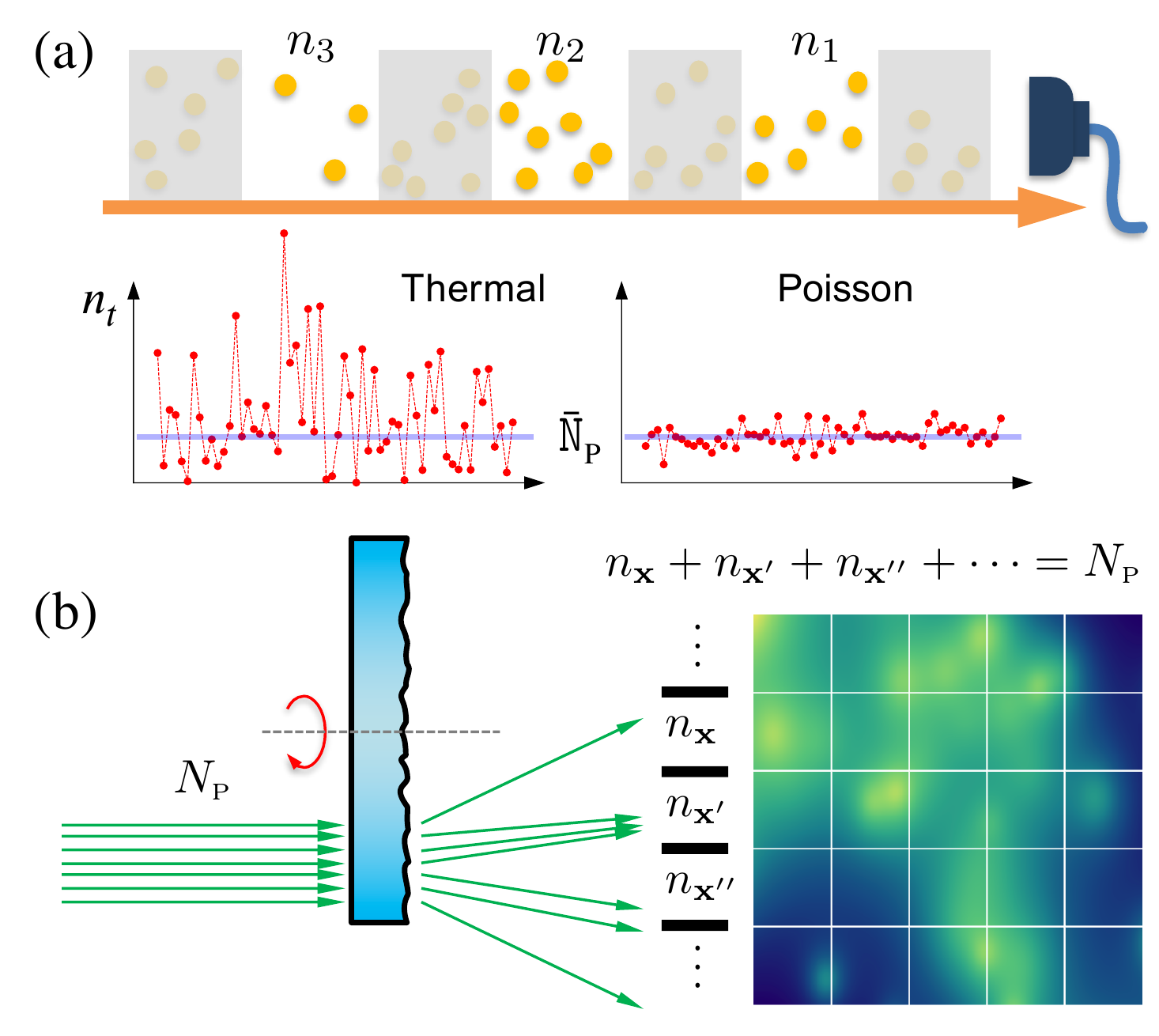}

\caption{(a) Demonstration for the photon counting process. (b) The light beam
is scattered divergently by the RGG and generates a random light pattern. }

\label{fig-glass}
\end{figure}

We first briefly review the basic photon counting process {[}Fig.\,\ref{fig-glass}(a){]}.
Considering a steady light flux coming into the photon detector (PD),
in each exposure period, a quantum projective measurement is made
on the light intensity. The exposure time must be much shorter than
the light coherence time, otherwise the measured result is indeed
a time average, which corresponds to classical treatments \citep{mandel_fluctuations_1959,mandel_fluctuations_1958}.
In the idealistic case, the data $\{n_{t}\}$ of many exposure frames
should be a stochastic series of integers due to the quantized feature
of photons, as well as the intrinsic randomness of quantum measurements
\citep{sperling_sub-binomial_2012,sperling_true_2012,jiang_photon-number-resolving_2007,kardynal_avalanchephotodiode-based_2008}.
This can be regarded as multiple repetitive measurements on the same
light state. Counting $\{n_{t}\}$ gives the photon number distribution
$P_{n}$ of the incoming light.

Therefore, the pseudo-thermal light generation can be understood as
follows. Considering a laser beam divergently scattered by a RGG,
we suppose all the scattered light is collected by a pixel lattice
composed of $M$($\gg1$) independent PDs. In each exposure, this
pixel lattice takes a photo frame of the whole scattered light, which
appears as a light pattern randomly distributed in spatial domain
{[}Fig.\,\ref{fig-glass}(b){]}. When the ground glass is moving,
this light pattern varies with time like a ``zoetrope''. Thus, when
focusing on a small area on the light pattern, its intensity fluctuates
with time stochastically, which is similar like the above quantum
measurement data. This is just how the pseudo-thermal light is obtained.

Thus, the photon statistics on a certain pixel can be calculated by
counting how the input photons are distributed on the whole pixel
lattice, as shown below.
\begin{flushleft}
\textbf{Fock input:}
\par\end{flushleft}

We first consider a simple case that in each frame, the incoming light
always contains exactly $\mathtt{N}_{\text{\textsc{p}}}$ photons
(Fock input). Due to the stochastic scattering, these $\mathtt{N}_{\text{\textsc{p}}}$
photons would be randomly distributed to all the $M$ pixels. Denoting
 $n_{\mathbf{x},t}$ as the photon number recorded at pixel-$\mathbf{x}$
in frame-$t$, all possible pattern configurations $\{n_{\mathbf{x}}\}$
may appear (assuming with equal probabilities). Remember all $n_{\mathbf{x}}$
are integers ($0\le n_{\mathbf{x}}\le\mathtt{N}_{\text{\textsc{p}}}$),
and they must satisfy the constraint of energy conservation $\sum_{\mathbf{x}}n_{\mathbf{x}}=\mathtt{N}_{\text{\textsc{p}}}$.
Thus, since photons are identical bosons,  the total number of such
combinations is given by \citep{landau_statistical_1980,huang_statistical_1987}\footnote{
This is equivalent as counting the combinations how $\mathtt{N}_{\text{\textsc{p}}}+M$
identical balls are put into $M$ different boxes, with zero reception
allowed.}
\begin{equation}
Z:=\binom{\mathtt{N}_{\text{\textsc{p}}}+M-1}{M-1}=\frac{(\mathtt{N}_{\text{\textsc{p}}}+M-1)!}{(M-1)!\,\mathtt{N}_{\text{\textsc{p}}}!}.
\end{equation}

Further, we count the probability that $n$ photons are received by
a single pixel. That is equivalent as counting the combinations how
$\mathtt{N}_{\text{\textsc{p}}}-n$ photons are distributed on the
rest $M-1$ pixels, which gives (for $M\ge2$)
\begin{equation}
P_{n}^{(\mathtt{N}_{\text{\textsc{p}}})}=\begin{cases}
\frac{1}{Z}\binom{\mathtt{N}_{\text{\textsc{p}}}-n+M-2}{M-2}, & 0\le n\le\mathtt{N}_{\text{\textsc{p}}}\\
0, & n>\mathtt{N}_{\text{\textsc{p}}}
\end{cases}\label{eq:Pn-0}
\end{equation}
This is just the scattered photon statistics on one pixel when the
input light carries $\mathtt{N}_{\text{\textsc{p}}}$ photons in each
frame.

\begin{figure}
\includegraphics[width=1\columnwidth]{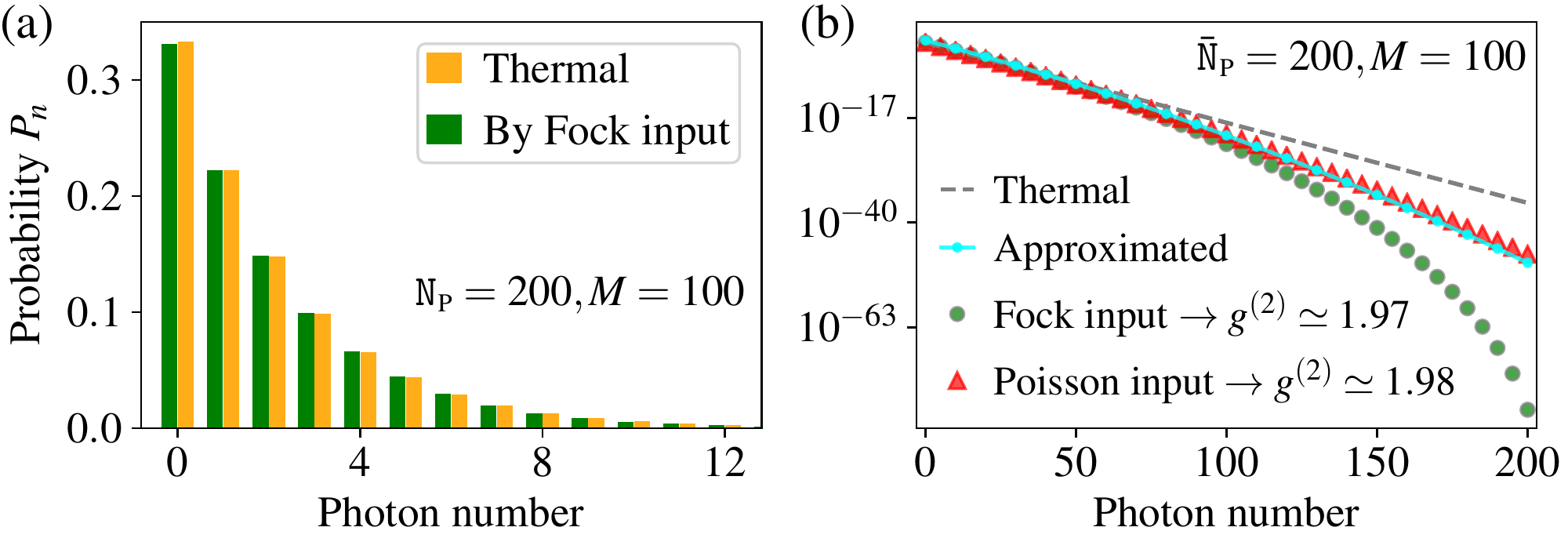}

\caption{(a) The scattered photon distribution $P_{n}$ on one pixel from Fock
input (left green), comparing with the thermal one (right yellow)
with the same mean photon number. (b) The scattered photon distribution
(in log scale) from the Fock and Poisson input ($\bar{\mathtt{N}}_{\text{\textsc{p}}}=200$)
comparing with the thermal distribution (dashed gray). The dotted
cyan line is the approximated result (\ref{eq:P_n-(Apx)}).}

\label{fig-Np}
\end{figure}

In the regime $n\ll\mathtt{N}_{\text{\textsc{p}}}$, the above probability
gives 
\begin{equation}
\frac{P_{n+1}^{(\mathtt{N}_{\text{\textsc{p}}})}}{P_{n}^{(\mathtt{N}_{\text{\textsc{p}}})}}=\frac{1}{1+\cfrac{M-2}{\mathtt{N}_{\text{\textsc{p}}}-n}}\simeq\frac{1}{1+M/\mathtt{N}_{\text{\textsc{p}}}},\label{eq:Thermal-0}
\end{equation}
 which is a constant and less than 1 (when $M\gg1$). That means,
the statistics received on this pixel well exhibits the thermal distribution.
In the regime $n\sim\mathtt{N}_{\text{\textsc{p}}}$, the deviation
from the ideal thermal distribution gradually grows larger, but this
difference is not easy to be observed in practical experiments (Fig.\,\ref{fig-Np}).

Therefore, when focusing on the photon statistics received on a local
area, the conservation constraint of the total photon number naturally
leads to the thermal statistics. Indeed this is quite similar with
the mechanism how the canonical ensemble with the thermal statistics
emerges as a subsystem inside a bigger micro-canonical one in statistical
physics \citep{huang_statistical_1987,landau_statistical_1980}.

\begin{flushleft}
\textbf{Poisson input:}
\par\end{flushleft}

The incoming light could also have certain distribution ${\cal P}_{\mathrm{in}}(\text{\textsc{n}})$
but not exactly $\mathtt{N}_{\text{\textsc{p}}}$ photons. In this
case, taking the input statistics into account, the scattered statistics
is 
\begin{equation}
P_{n}=\sum_{\text{\textsc{n}}}{\cal P}_{\mathrm{in}}(\text{\textsc{n}})\cdot P_{n}^{(\text{\textsc{n}})},\label{eq:Pn-Laser}
\end{equation}
where $P_{n}^{(\text{\textsc{n}})}$ is the result from the scattering
of $\text{\textsc{n}}$ photons {[}Eq.\,(\ref{eq:Pn-0}){]}.

For a laser input, the input statistics ${\cal P}_{\mathrm{in}}(\text{\textsc{n}})$
is approximately a Poisson distribution, which is narrowly distributed
around its mean value $\bar{\mathtt{N}}_{\text{\textsc{p}}}$ with
a relatively small variance ($\sqrt{\langle\delta n^{2}\rangle}\big/\langle n\rangle=1/\sqrt{\bar{\mathtt{N}}_{\text{\textsc{p}}}}$).
Therefore, in the above summation (\ref{eq:Pn-Laser}), only the terms
around $\text{\textsc{n}}\simeq\bar{\mathtt{N}}_{\text{\textsc{p}}}$
contribute significantly. As a result, approximately the output photon
statistics is also a thermal one, which satisfies 
\begin{equation}
\frac{P_{n+1}}{P_{n}}\simeq\frac{1}{1+M/\bar{\mathtt{N}}_{\text{\textsc{p}}}}.\label{eq:Pn-T}
\end{equation}

For the above Fock and Poisson input cases, the exact output distributions
$P_{n}$ are numerically shown in Fig.\,\ref{fig-Np}(a, b), and
they are both quite close to the thermal distribution in the regime
$n\ll\bar{\mathtt{N}}_{\text{\textsc{p}}}$. When the photon number
$n$ is large, both of them deviate from the thermal one.

\begin{flushleft}
\textbf{Non-thermal correction:}
\par\end{flushleft}

Here we give an approximated distribution taking account the above
non-thermal correction. We expand the above iterative relation (\ref{eq:Thermal-0})
as follows 
\begin{align}
\ln P_{n+1}^{(\mathtt{N}_{\text{\textsc{p}}})} & -\ln P_{n}^{(\mathtt{N}_{\text{\textsc{p}}})}=\ln\frac{1}{1+\frac{M-2}{\mathtt{N}_{\text{\textsc{p}}}-n}}\nonumber \\
\simeq & \ln\frac{1}{1+\frac{M-2}{\mathtt{N}_{\text{\textsc{p}}}}}-n\cdot\frac{M-2}{\mathtt{N}_{\text{\textsc{p}}}(\mathtt{N}_{\text{\textsc{p}}}+M-2)}+o(n^{2}).\label{eq:expansion}
\end{align}
Omitting the high order terms $o(n^{2})$, the summation over $n$
on both sides gives the following approximated distribution {[}the
dotted cyan line in Fig.\,\ref{fig-Np}(b){]}
\begin{align}
P_{n}^{(\mathtt{N}_{\text{\textsc{p}}})} & =P_{0}e^{-\beta_{0}n-\beta_{\text{c}}(n^{2}-n)},\label{eq:P_n-(Apx)}\\
\beta_{0} & =\ln\big(1+\frac{M-2}{\mathtt{N}_{\text{\textsc{p}}}}\big),\quad\beta_{\text{c}}=\frac{1}{2}\frac{M-2}{\mathtt{N}_{\text{\textsc{p}}}(\mathtt{N}_{\text{\textsc{p}}}+M-2)},\nonumber 
\end{align}
where $P_{0}$ is a normalization constant. A quadratic correction
appears in the exponential factor, and the correction $\beta_{\text{c}}$
is negligible for strong input $\mathtt{N}_{\text{\textsc{p}}}$.
Indeed, this is a typical non-canonical feature resulted from finite
system sizes or coupling strengths \citep{xu_noncanonical_2014,dong_quantum_2007}.

At last we remark that the ``pixels'' in the above discussions do
not corresponds to realistic detector pixels directly. The sizes of
such pixels are determined by the spatial correlation length of the
scattered light pattern, so they can be regarded as independent fluctuation
units. And their total number $M$ is roughly determined by the granules
within the light spot on the RGG, which is usually a very large number
in practice. 

\section{The scattered statistics from non-classical light }

If the input light is a non-classical state, the standard classical
theory does not apply \citep{goodman_speckle_2010,goodman_statistical_2000}.
In this case, generally the scattered photon statistics is no longer
a thermal one. As long as the corresponding input $\mathcal{P}_{\text{in}}(\text{\textsc{n}})$
is taken, the general result for the scattered photon statistics $P_{n}$
can be obtained {[}Eq.\,(\ref{eq:Pn-Laser}){]}.

\begin{figure}
\includegraphics[width=1\columnwidth]{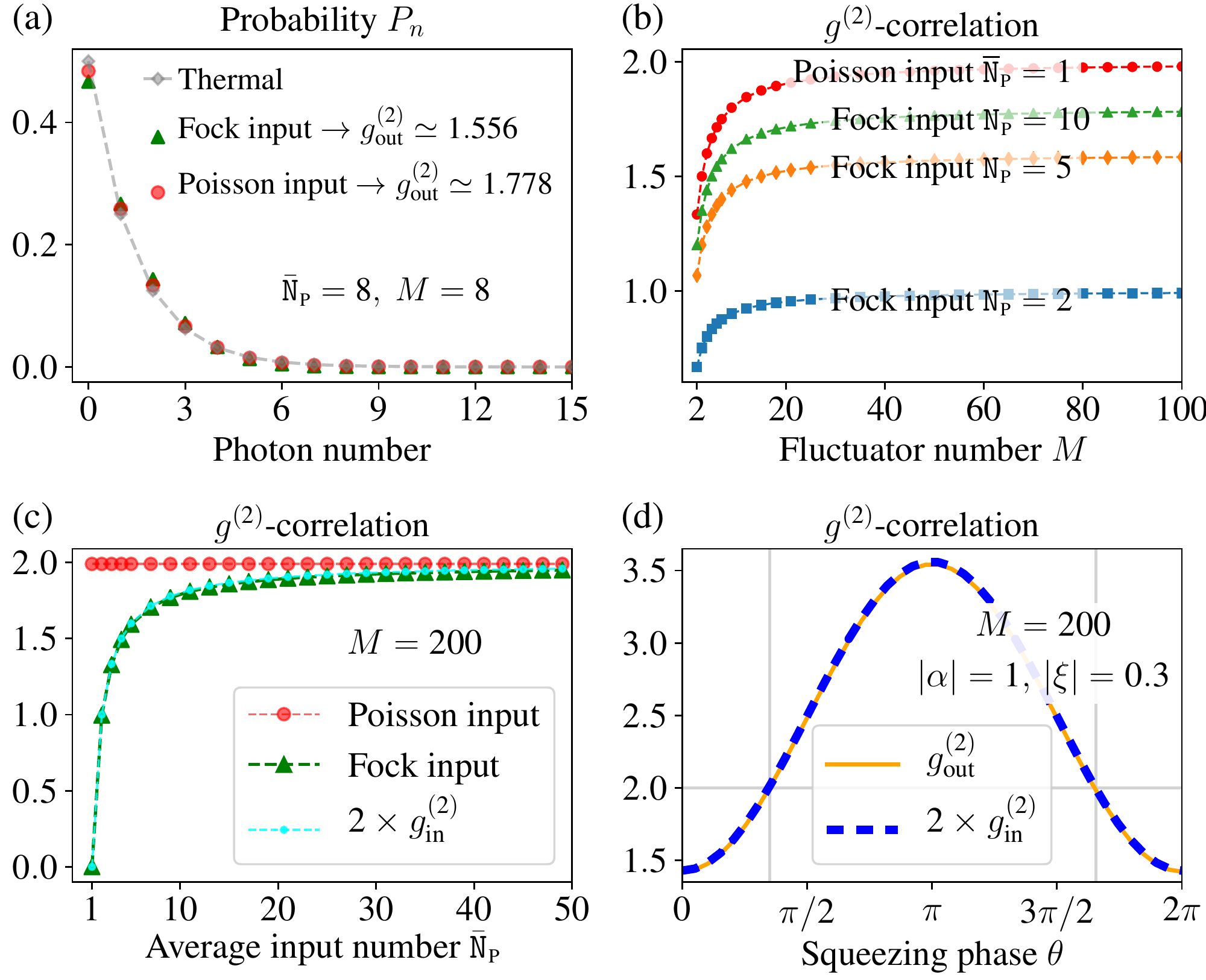}

\caption{(a) The scattered distribution $P_{n}$ from Fock and Poisson input
($\bar{\mathtt{N}}_{\text{\textsc{p}}}=8$, $M=8$). (b) The scattered
$g_{\text{out}}^{(2)}$ from Fock ($\mathtt{N}_{\text{\textsc{p}}}=2,5,10$)
and Poisson input ($\bar{\mathtt{N}}_{\text{\textsc{p}}}=1$) changing
with fluctuator number $M$. (c) The scattered $g_{\text{out}}^{(2)}$
changing with the input photon number (given $M=200$), comparing
with the input $g_{\text{in}}^{(2)}$ of Fock state. (d) The scattered
$g_{\text{out}}^{(2)}$ from squeezed input changing with the squeezing
phase $\theta$, comparing with the input $g_{\text{in}}^{(2)}$.}

\label{fig-Np-2}
\end{figure}

Comparing with the distribution profile $P_{n}$, the $g^{(2)}$-correlation
is indeed a more precise indicator to characterize the photon statistics
\citep{agarwal_quantum_2012,scully_quantum_1997}. For example, considering
averagely $\bar{\mathtt{N}}_{\text{\textsc{p}}}=8$ photons scattered
onto $M=8$ fluctuating units, although the scattered $P_{n}$ from
both Fock and Poisson inputs look quite close to the thermal one {[}Fig.\,\ref{fig-Np-2}(a){]},
indeed they give $g_{\text{out}}^{(2)}\simeq1.556$ (Fock input) and
$g_{\text{out}}^{(2)}\simeq1.778$ (Poisson input), which are both
significantly different from the thermal result $g_{\text{th}}^{(2)}=2$. 

For arbitrary input states, the scattered $g^{(2)}$-correlation can
be calculated from Eq.\,(\ref{eq:Pn-Laser}). The mean photon number
of the scattered light is $\langle n\rangle=\langle\text{\textsc{n}}\rangle_{\text{in}}/M$,
and the mean square is: 
\begin{align}
\langle n^{2}\rangle & =\sum_{\text{\textsc{n}},\,n}\mathcal{P}_{\text{in}}(\text{\textsc{n}})\cdot n^{2}P_{n}^{(\text{\textsc{n}})}=\sum_{\text{\textsc{n}}}\mathcal{P}_{\text{in}}(\text{\textsc{n}})\frac{\text{\textsc{n}}(2\text{\textsc{n}}+M-1)}{M(M+1)}\nonumber \\
 & =\frac{2\langle\text{\textsc{n}}^{2}\rangle_{\text{in}}}{M(M+1)}+\frac{\langle\text{\textsc{n}}\rangle_{\text{in}}(M-1)}{M(M+1)}.\label{eq:n2}
\end{align}
Then after simple calculations, the scattered $g_{\text{out}}^{(2)}=\big(\langle n^{2}\rangle-\langle n\rangle\big)/\langle n\rangle^{2}$
can be represented by the input $g_{\text{in}}^{(2)}$:
\begin{equation}
g_{\text{out}}^{(2)}=\frac{2\big(\langle\text{\textsc{n}}^{2}\rangle_{\text{in}}-\langle\text{\textsc{n}}\rangle_{\text{in}}\big)M}{\langle\text{\textsc{n}}\rangle_{\text{in}}^{2}(M+1)}=2g_{\text{in}}^{(2)}\cdot\frac{M}{M+1}.\label{eq:g-out-g-in}
\end{equation}

In practice, the RGG usually has a large number of diffusers ($M\gg1$),
which gives $g_{\text{out}}^{(2)}\simeq2g_{\text{in}}^{(2)}$. It
is worth noticing this is a general result for arbitrary input statistics,
including non-classical light.

Therefore, a laser input with Poisson statistics always leads to the
thermal result $g_{\text{out}}^{(2)}\simeq2$, which is irrelevant
of the input intensity, even at the single-photon level $\bar{\mathtt{N}}_{\text{\textsc{p}}}=1$. 

On contrary, the Fock input $|\mathtt{N}_{\text{\textsc{p}}}\rangle$
has $g_{\text{in}}^{(2)}=1-\mathtt{N}_{\text{\textsc{p}}}^{-1}\in[0,1)$,
and that makes the scattered $g_{\text{out}}^{(2)}\simeq2g_{\text{in}}^{(2)}$
increase with the input photon number $\mathtt{N}_{\text{\textsc{p}}}$
{[}Fig.\,\ref{fig-Np-2}(c){]}. In particular, the single-photon
input leads to a scattered state $\hat{\rho}_{\text{out}}=(1-\frac{1}{M})|0\rangle\langle0|+\frac{1}{M}|1\rangle\langle1|$
with $g_{\text{out}}^{(2)}=0$.

Thus, the quantum features of non-classical states appear more significant
in the few-photon regime. And it is also worth to notice that the
Poisson input plays better in producing the thermal statistics than
the Fock input, since it always guarantees $g_{\text{out}}^{(2)}\simeq2$
irrelevant of the input intensity.

With the increase of the total number $M$ of the fluctuating units,
the scattered $g_{\text{out}}^{(2)}$ increases and converges to $g_{\text{out}}^{(2)}\simeq2g_{\text{in}}^{(2)}$
{[}Fig.\,\ref{fig-Np-2}(b){]}. And this just corresponds to the
condition of infinite diffusers required by the central limit theorem
in classical theory. 

By utilizing the squeezed light, a non-classical input state with
sub-Poisson statistics can be realized in experiments \citep{liu_enhanced_2016,agarwal_quantum_2012,gerry_introductory_2005}.
In Fig.\,\ref{fig-Np-2}(d) we show the scattered $g_{\text{out}}^{(2)}$
from the squeezed light calculated from Eq.\,(\ref{eq:Pn-Laser}).
The input statistics $\mathcal{P}_{\text{in}}(\text{\textsc{n}})$
comes from the squeezed state $|\alpha,\xi\rangle=\hat{D}_{\alpha}\hat{S}_{\xi}|0\rangle$,
where $\hat{S}_{\xi}:=\exp\big[\frac{1}{2}(\xi^{*}\hat{a}^{2}-\xi\hat{a}^{\dagger}{}^{2})\big]$,
$\hat{D}_{\alpha}:=\exp[\alpha\hat{a}^{\dagger}-\alpha^{*}\hat{a}]$
are the squeezing and displacement operators, with $\xi=re^{i\theta}$
($r\ge0$) as the squeezing parameter \citep{gerry_introductory_2005,agarwal_quantum_2012}. 

For different squeezing phases $\theta$, the mean photon number keeps
to be $\bar{\mathtt{N}}_{\text{\textsc{p}}}=|\alpha|^{2}+\sinh^{2}r$,
while the input statistics gives sub-Poisson ($g_{\text{in}}^{(2)}<1$)
or super-Poisson ($g_{\text{in}}^{(2)}>1$) distributions. In the
whole regime the above relation $g_{\text{out}}^{(2)}\simeq2g_{\text{in}}^{(2)}$
well applies.

Notice that the output light can be used as the input light for further
scattering through another RGG, then that gives $g_{\text{out-2}}^{(2)}\simeq2g_{\text{out-1}}^{(2)}\simeq2^{2}g_{\text{in}}^{(2)}$.
Clearly, after the hierarchy scattering through $k$ RGGs, the final
output light has $g_{\text{out-}k}^{(2)}\simeq2^{k}g_{\text{in}}^{(2)}$.
In particular, if the original input is laser, the final output gives
$g_{\text{out-}k}^{(2)}\simeq2^{k}$ \citep{goodman_statistical_2000,zhou_superbunching_2017,goodman_speckle_2010}.

Besides, similarly like Eq.\,(\ref{eq:n2}), the 3rd order expectation
$\langle n^{3}\rangle$ and $g_{\text{out}}^{(3)}$ also can be obtained,
i.e.,
\begin{align}
\langle n^{3}\rangle & =\frac{6\langle\text{\textsc{n}}^{3}\rangle_{\text{in}}+6(M-1)\langle\text{\textsc{n}}^{2}\rangle_{\text{in}}+(M^{2}-3M+2)\langle\text{\textsc{n}}\rangle_{\text{in}}}{M(M+1)(M+2)},\nonumber \\
g_{\text{out}}^{(3)} & =\frac{\langle\hat{a}^{\dagger3}\hat{a}^{3}\rangle_{\text{out}}}{\langle\hat{a}^{\dagger}\hat{a}\rangle_{\text{out}}^{3}}=6g_{\text{in}}^{(3)}\cdot\frac{M^{2}}{M^{2}+3M+2},
\end{align}
which gives $g_{\text{out}}^{(3)}\simeq6g_{\text{in}}^{(3)}$ for
large $M$. But the calculation of higher orders becomes more and
more troublesome. In the following, with the help of the \emph{P}-function
of the scattered light, we can prove $g_{\text{out}}^{(n)}\simeq n!\,g_{\text{in}}^{(n)}$
for any order-$n$ and any input state.

\section{The \emph{P}-function of the scattered light and the high-order correlations}

The non-classical properties of a light state can be more clearly
characterized by its \emph{P}-function \citep{scully_quantum_1997,agarwal_quantum_2012,agarwal_nonclassical_1992},
and here we show the \emph{P}-function of the scattered light and
its high-order correlations. As seen above, the Poisson input plays
better in producing the thermal statistics than the Fock input, since
it always leads to the thermal result $g_{\text{out}}^{(2)}\simeq2$
in spite of the input intensity (in the many-diffuser limit $M\rightarrow\infty$).
Thus, approximately this can be written as the following mapping relation:
\begin{alignat}{2}
\hat{\rho}_{\text{in}} & =|\alpha_{0}\rangle\langle\alpha_{0}| & \;\rightarrow\; & \hat{\rho}_{\text{out}}=\sum\frac{\bar{\text{\textsc{n}}}_{T}{}^{m}}{(1+\bar{\text{\textsc{n}}}_{T})^{n+1}}|n\rangle\langle n|,\nonumber \\
P_{\text{in}}(\alpha) & =\delta^{(2)}(\alpha-\alpha_{0}) & \;\rightarrow\; & P_{\text{out}}(\alpha)=\frac{1}{\pi\bar{\text{\textsc{n}}}_{T}}\exp\big(-\frac{|\alpha|^{2}}{\bar{\text{\textsc{n}}}_{T}}\big).\label{eq:map}
\end{alignat}
 Here $|\alpha_{0}|^{2}$ and $\bar{\text{\textsc{n}}}_{T}=|\alpha_{0}|^{2}/M$
are the mean photon numbers of the input and output statistics.

This input-output relation can be written as a linear functional transformation
$P_{\text{out}}(\alpha)=\mathscr{F}_{\alpha}\big[P_{\text{in}}(\alpha)\big]$.
Then for a general input state $P_{\text{in}}(\alpha)=\int d^{2}\varsigma\,P_{\text{in}}(\varsigma)\delta^{(2)}(\alpha-\varsigma)$,
the output \emph{P}-function is 
\begin{align}
P_{\text{out}}(\alpha) & =\mathscr{F}_{\alpha}\big[P_{\text{in}}(\alpha)\big]=\int d^{2}\varsigma\,P_{\text{in}}(\varsigma)\mathscr{F}_{\alpha}[\delta^{(2)}(\alpha-\varsigma)]\nonumber \\
 & =\int d^{2}\varsigma\,P_{\text{in}}(\varsigma)\cdot\frac{1}{\pi|\varsigma|^{2}/M}\exp\big(-\frac{|\alpha|^{2}}{|\varsigma|^{2}/M}\big).\label{eq:P-out}
\end{align}

First, we use Eq.\,(\ref{eq:P-out}) to obtain the high-order moments
of the scattered light,
\begin{align}
\langle\hat{a}^{\dagger n} & \hat{a}^{n}\rangle_{\text{out}}=\int d^{2}\alpha\,|\alpha|^{2n}P_{\text{out}}(\alpha)\nonumber \\
= & \int d^{2}\varsigma\int d^{2}\alpha\,\frac{M}{\pi|\varsigma|^{2}}e^{-\frac{M}{|\varsigma|^{2}}|\alpha|^{2}}|\alpha|^{2n}\cdot P_{\text{in}}(\varsigma)\nonumber \\
= & \int d^{2}\varsigma\,P_{\text{in}}(\varsigma)\cdot\frac{|\varsigma|^{2n}}{M^{n}}n!=\langle\hat{a}^{\dagger n}\hat{a}^{n}\rangle_{\text{in}}\cdot\frac{n!}{M^{n}}.
\end{align}
Therefore, for arbitrary input states, the $n$-order correlation
function of the scattered light is 
\begin{equation}
g_{\text{out}}^{(n)}=\frac{\langle\hat{a}^{\dagger n}\hat{a}^{n}\rangle_{\text{out}}}{\big(\langle\hat{a}^{\dagger}\hat{a}\rangle_{\text{out}}\big)^{n}}=\frac{n!\,\langle\hat{a}^{\dagger n}\hat{a}^{n}\rangle_{\text{in}}/M^{n}}{\big(\langle\hat{a}^{\dagger}\hat{a}\rangle_{\text{in}}/M\big)^{n}}=n!\,g_{\text{in}}^{(n)}.
\end{equation}
This result holds for arbitrary input statistics. In particular, the
2nd order gives $g_{\text{out}}^{(2)}=2g_{\text{in}}^{(2)}$, which
is just the result (\ref{eq:g-out-g-in}) in the many-diffuser limit
$M\rightarrow\infty$. After the hierarchy scattering through $k$
RGGs, the final output light gives $g_{\text{out-}k}^{(n)}\simeq(n!)^{k}g_{\text{in}}^{(n)}$.

Next we use the output \emph{P}-function (\ref{eq:P-out}) to obtain
the photon statistics, and compare it with the above result (\ref{eq:Pn-0})
based on combination counting. Since the scattered state can be written
as $\hat{\rho}_{\text{out}}=\int d^{2}\alpha\,P_{\text{out}}(\alpha)|\alpha\rangle\langle\alpha|=\sum\tilde{P}_{n}|n\rangle\langle n|$,
the photon number distribution $\tilde{P}_{n}$ also can be obtained
as 
\begin{align}
\tilde{P}_{n} & =\langle n|\hat{\rho}_{\text{out}}|n\rangle=\int d^{2}\alpha\,P_{\text{out}}(\alpha)\cdot\frac{|\alpha|^{2n}}{n!}e^{-|\alpha|^{2}}\nonumber \\
 & =\int d^{2}\varsigma\,P_{\text{in}}(\varsigma)\,\frac{M/|\varsigma|^{2}}{(1+M/|\varsigma|^{2})^{n+1}}.\label{eq:Pn-2int}
\end{align}

For a Fock input state $|\mathtt{N}\rangle$, the input \emph{P}-function
is $P_{\text{in}}(\varsigma)=e^{|\varsigma|^{2}}[\partial_{\varsigma}^{\mathtt{N}}\partial_{\varsigma^{*}}^{\mathtt{N}}\delta^{(2)}(\varsigma)]\big/\mathtt{N}!$
\citep{agarwal_quantum_2012}, and the above integral gives 
\begin{align}
\tilde{P}_{n}^{(\mathtt{N})} & =\lim_{\varsigma\rightarrow0}\,\frac{1}{\mathtt{N}!}\partial_{\varsigma}^{\mathtt{N}}\partial_{\varsigma^{*}}^{\mathtt{N}}\Big[\frac{e^{|\varsigma|^{2}}\,M/|\varsigma|^{2}}{(1+M/|\varsigma|^{2})^{n+1}}\Big]\nonumber \\
 & =\frac{\mathtt{N}!}{n!}\sum_{k=n}^{\mathtt{N}}\frac{(-1)^{k-n}k!}{(\mathtt{N}-k)!(k-n)!M^{k}},\label{eq:Pn-2deriv}
\end{align}
and $\tilde{P}_{n}^{(\mathtt{N})}=0$ for $n>\mathtt{N}$ (see Appendix
\ref{Apx:Derivative}). 

\begin{figure}
\includegraphics[width=1\columnwidth]{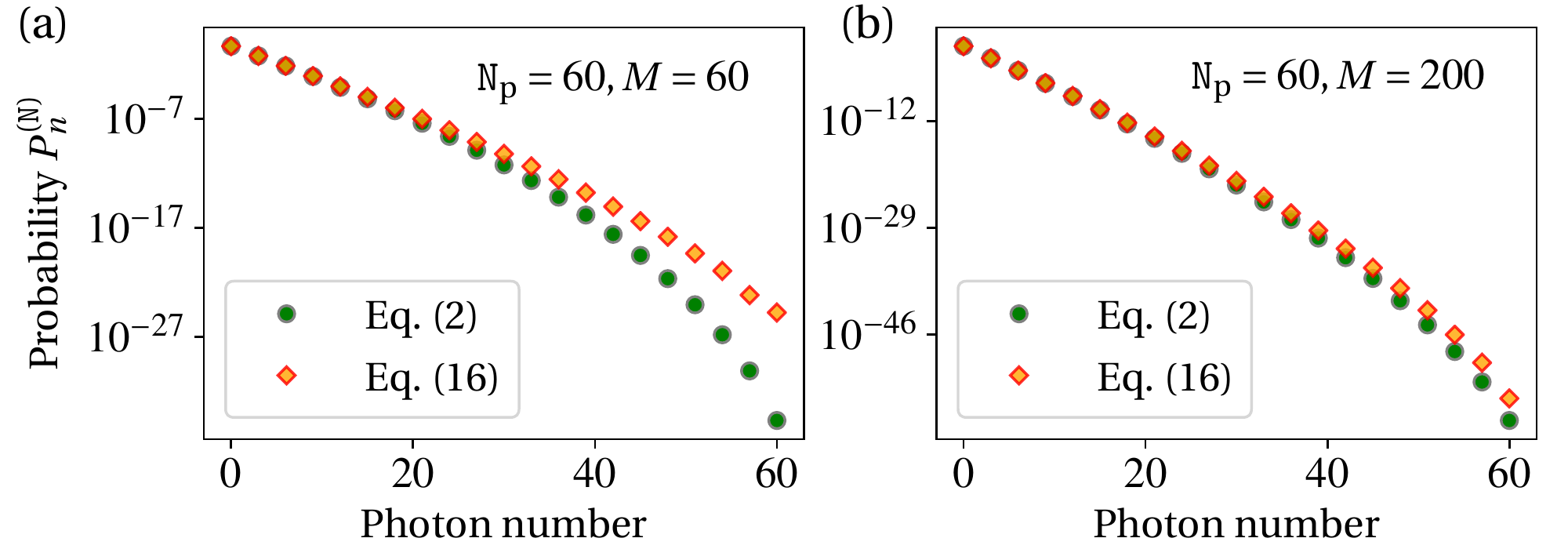}

\caption{The scattered distribution $P_{n}$ from Fock input ($\mathtt{N}_{\text{\textsc{p}}}=60$)
calculated from Eq.\,(\ref{eq:Pn-0}) (based on combination counting)
and Eq.\,(\ref{eq:Pn-2deriv}) {[}based on the mapping relation (\ref{eq:map}){]}
for (a) $M=60$ (b) $M=200$. For large $M$, they almost coincide
with each other.}

\label{fig-Pn1Pn2}
\end{figure}

Here we emphasize Eqs.\,(\ref{eq:Pn-2int}, \ref{eq:Pn-2deriv})
are valid only in the many-diffuser limit $M\rightarrow\infty$, which
guarantees the Poisson input must exactly produce the thermal state
and $g_{\text{out}}^{(2)}=2g_{\text{in}}^{(2)}$ holds exactly. Thus
generally $\tilde{P}_{n}^{(\mathtt{N})}$ is not equivalent with the
result (\ref{eq:Pn-0}) based on combination counting, which applies
in more general cases. When $M$ is large, their difference becomes
negligible, especially in the few-photon regime (Fig.\,\ref{fig-Pn1Pn2}). 

\section{Summary}

In this paper, we develop a quantum framework to study the full photon
number statistics of the scattered light passing through the RGG.
The output statistics is obtained by counting the combinations how
the input photons are distributed on the scattered light pattern.
When the total photon number of the whole scattered light pattern
is approximately a constant, counting the photon number received by
a small area on the light pattern naturally gives the thermal statistics.
Then for arbitrary input states, the scattered photon statistics also
can be well obtained by taking the input statistics into account.

We also obtain the \emph{P}-function of the scattered light from arbitrary
input states in the many-diffuser limit. With the help of these distributions,
we find that the scattered light always gives a relation $g_{\text{out}}^{(n)}\simeq n!\,g_{\text{in}}^{(n)}$,
which holds for any order-$n$ and any input state, including non-classical
ones. In these situations, this theory provides a more precise description
beyond the previous classical theory based on central limit theorem. 

The above results of $g^{(n)}$-correlations scattered from non-classical
input can be verified by squeezed light in current experiments, and
may be utilized for correlation imaging of high orders. This result
also indicates it is possible to create a scattered light with very
high $g^{(n)}$-correlations simply, which may be utilized to enhance
the multi-photon absorption processes and high harmonic generation
\citep{spasibko_multiphoton_2017,agarwal_field-correlation_1970,jechow_enhanced_2013}.

\vspace{.6em}

\emph{Acknowledgment} -- SWL appreciates very much for the helpful
discussion with R. Nessler in Texas A\&M University, W.-K. Yu in Beijing
Institute of Technology, and J. Sperling in Universit\"at Rostock.
This study is supported by NSF of China (Grant No.\,11905007), Beijing
Institute of Technology Research Fund Program for Young Scholars,
AFOSR award FA-9550-18-1-0141, ONR award N00014-16-1-3054, and Robert
A. Welch Foundation award A-1261.

\appendix
\begin{widetext}

\section{Calculation of Eq.\,(\ref{eq:Pn-2deriv}) \label{Apx:Derivative}}

The \emph{P}-function of the Fock state $|\mathtt{N}\rangle$ is a
highly singular function $P_{\text{in}}(\varsigma)=e^{|\varsigma|^{2}}[\partial_{\varsigma}^{\mathtt{N}}\partial_{\varsigma^{*}}^{\mathtt{N}}\delta^{(2)}(\varsigma)]\big/\mathtt{N}!$,
which contains the derivative of $\delta$-function. Notice that for
$\varsigma=\varsigma_{x}+i\varsigma_{y}$, the derivative gives $\partial_{\varsigma}\partial_{\varsigma^{*}}=\frac{1}{4}(\partial_{\varsigma_{x}}^{2}+\partial_{\varsigma_{y}}^{2})=\frac{1}{4}\nabla^{2}$.
From the Stokes theorem we obtain $\int d^{2}\varsigma\,\big[f(\varsigma)\nabla^{2n}\delta^{(2)}(\varsigma)-\delta^{(2)}(\varsigma)\nabla^{2n}f(\varsigma)\big]=0$,
thus the integral (\ref{eq:Pn-2int}) gives 
\begin{align}
\tilde{P}_{n} & =\frac{1}{\mathtt{N}!}\int d^{2}\varsigma\,\big[\frac{\partial^{2\mathtt{N}}}{\partial\varsigma^{\mathtt{N}}\partial\varsigma^{*\mathtt{N}}}\delta^{(2)}(\varsigma)\big]\,e^{|\varsigma|^{2}}\frac{M/|\varsigma|^{2}}{(1+M/|\varsigma|^{2})^{n+1}}=\frac{1}{\mathtt{N}!}\int d^{2}\varsigma\,\delta^{(2)}(\varsigma)\,\frac{\partial^{2\mathtt{N}}}{\partial\varsigma^{\mathtt{N}}\partial\varsigma^{*\mathtt{N}}}\frac{e^{|\varsigma|^{2}}M/|\varsigma|^{2}}{(1+M/|\varsigma|^{2})^{n+1}}\nonumber \\
 & =\lim_{\varsigma\rightarrow0}\,\frac{1}{\mathtt{N}!}\partial_{\varsigma}^{\mathtt{N}}\partial_{\varsigma^{*}}^{\mathtt{N}}\Big[\frac{e^{|\varsigma|^{2}}\,M/|\varsigma|^{2}}{(1+M/|\varsigma|^{2})^{n+1}}\Big].
\end{align}

Now we need to calculate this derivative and taking the limit. Denoting
$\varsigma\rightarrow x$, $\varsigma^{*}\rightarrow y$, the above
derivative gives 

\begin{align}
\partial_{y}^{\mathtt{N}}\Big[\frac{e^{xy}\,M(xy)^{n}}{(M+xy)^{n+1}}\Big] & =\sum_{k=0}^{\mathtt{N}}C_{\mathtt{N}}^{k}\cdot\partial_{y}^{\mathtt{N}-k}e^{xy}\cdot\partial_{y}^{k}\frac{M(xy)^{n}}{(M+xy)^{n+1}}=\sum_{k=0}^{\mathtt{N}}C_{\mathtt{N}}^{k}\cdot e^{xy}x^{\mathtt{N}-k}\cdot\sum_{q=0}^{k}C_{k}^{q}\cdot\partial_{y}^{q}[M(xy)^{n}]\cdot\partial_{y}^{k-q}\frac{1}{[M+xy]^{n+1}}\nonumber \\
 & =\sum_{k=0}^{\mathtt{N}}\sum_{q=0}^{k}C_{\mathtt{N}}^{k}C_{k}^{q}\cdot e^{xy}x^{\mathtt{N}-k}\cdot Mx^{n}\frac{n!}{(n-q)!}y^{n-q}\cdot(-1)^{k-q}\frac{(n+k-q)!}{n!}(M+xy)^{-n-1-k+q}x^{k-q}\nonumber \\
 & =M\sum_{k=0}^{\mathtt{N}}\sum_{q=0}^{k}\frac{(-1)^{k-q}(n+k-q)!}{(n-q)!}C_{\mathtt{N}}^{k}C_{k}^{q}\left[y^{n-q}\cdot x^{\mathtt{N}-q+n}e^{xy}(M+xy)^{-n-1-k+q}\right].\label{eq:derivative}
\end{align}
Now we need to further calculate the derivative $\partial_{x}^{\mathtt{N}}$
by applying the Leibniz rule to the above bracket, which gives $\sum_{l=0}^{\mathtt{N}}C_{\mathtt{N}}^{l}\cdot[\partial_{x}^{l}x^{\mathtt{N}-q+n}]\cdot\partial_{x}^{\mathtt{N}-l}[e^{xy}(M+xy)^{-n-1-k+q}]$. 

Since at last we need to take the limit $x,y\rightarrow0$, from the
power factor $\partial_{x}^{l}x^{\mathtt{N}-q+n}=\frac{(\mathtt{N}-q+n)!}{(\mathtt{N}-q+n-l)!}x^{\mathtt{N}-q+n-l}$
we can notice that only few terms in the above summations could exist,
and they must satisfy $\mathtt{N}-q+n-l=0$, thus we have $q-n=\mathtt{N}-l\ge0$.
But since the denominator of Eq.\,(\ref{eq:derivative}) contains
$(n-q)!$ which gives $+\infty$ for $n<q$, that guarantees $q=n$
and $l=\mathtt{N}$. Therefore, taking the limit of the above derivative
gives
\begin{equation}
\lim_{x,y\rightarrow0}\partial_{x}^{\mathtt{N}}\partial_{y}^{\mathtt{N}}\Big[\frac{e^{xy}\,M(xy)^{n}}{(M+xy)^{n+1}}\Big]=\sum_{k=0}^{\mathtt{N}}(-1)^{k-n}k!\cdot\frac{\mathtt{N}!}{(\mathtt{N}-k)!k!}\cdot\frac{k!}{(k-n)!n!}\cdot\frac{\mathtt{N}!}{M^{k}},
\end{equation}
 as shown in the main text. 

\end{widetext}

\bibliographystyle{apsrev4-1}
\bibliography{Refs}

\end{document}